\documentclass[11pt]{article}
\usepackage{amsfonts,amsmath,amssymb,amsfonts,graphicx,hyperref}
\input{tcilatex}

\begin{document}

\title{Adaptive--Wave Alternative for\\
the Black--Scholes Option Pricing Model}
\author{Vladimir G. Ivancevic \\
{\small Defence Science \& Technology Organisation, Australia}}
\date{}
\maketitle

\begin{abstract}
A nonlinear wave alternative for the standard Black--Scholes option--pricing model is presented. The adaptive-wave model, representing \emph{controlled Brownian behavior} of financial markets, is formally defined by adaptive nonlinear Schr\"odinger (NLS) equations, defining the option-pricing wave function in terms of the stock price and time. The model includes two parameters: volatility (playing the role of dispersion frequency coefficient), which can be either fixed or stochastic, and adaptive market potential that depends on the interest rate. The wave function represents quantum probability amplitude, whose absolute square is probability density function. Four types of analytical solutions of the NLS equation are provided in terms of Jacobi elliptic functions, all starting from de Broglie's plane-wave packet associated with the free quantum-mechanical particle. The best agreement with the Black--Scholes model shows the adaptive shock-wave NLS-solution, which can be efficiently combined with adaptive solitary-wave NLS-solution. Adjustable `weights' of the adaptive market-heat potential are estimated using either unsupervised Hebbian learning, or supervised Levenberg--Marquardt algorithm. In the case of stochastic volatility, it is itself represented by the wave function, so we come to the so-called Manakov system of two coupled NLS equations (that admits closed-form solutions), with the common adaptive market potential, which defines a bidirectional spatio-temporal associative memory.\\

\noindent\textbf{Keywords:} Black--Scholes option pricing, adaptive nonlinear Schr\"odinger equation,\\ market heat potential, controlled stochastic volatility, adaptive Manakov system,\\ controlled Brownian behavior
\end{abstract}


\newpage

\section{Introduction}

The celebrated Black--Scholes partial differential equation (PDE) describes
the time--evolution of the market value of a \textit{stock option} \cite%
{BS,Merton}. Formally, for a function $u=u(t,s)$ defined on the domain $%
0\leq s<\infty ,~0\leq t\leq T$ and describing the market value of a stock
option with the stock (asset) price $s$, the \emph{Black--Scholes PDE} can
be written (using the physicist notation: $\partial _{z}u=\partial
u/\partial z$) as a diffusion--type equation:
\begin{equation}
\partial _{t}u=-\frac{1}{2}(\sigma s)^{2}\,\partial _{ss}u-rs\,\partial
_{s}u+ru,  \label{BS}
\end{equation}
where $\sigma >0$ is the standard deviation, or \emph{volatility} of $s$, $r$
is the short--term prevailing continuously--compounded risk--free interest
rate, and $T>0$ is the time to maturity of the stock option. In this
formulation it is assumed that the \emph{underlying} (typically the stock)
follows a \emph{geometric Brownian motion} with `drift' $\mu $ and
volatility $\sigma$, given by the stochastic differential equation (SDE)
\cite{Osborne}
\begin{equation}
ds(t)=\mu s(t)dt+\sigma s(t)dW(t),  \label{gbm}
\end{equation}
where $W$ is the standard Wiener process.

The economic ideas behind the Black--Scholes option pricing theory
translated to the stochastic methods and concepts are as follows (see \cite%
{Perello}). First, the option price depends on the stock price and this is a
random variable evolving with time. Second, the efficient market hypothesis
\cite{Fama,Jensen}, i.e., the market incorporates instantaneously any
information concerning future market evolution, implies that the random term
in the stochastic equation must be delta--correlated. That is: speculative
prices are driven by white noise. It is known that any white noise can be
written as a combination of the derivative of the Wiener process \cite%
{Wiener} and white shot noise (see \cite{Gardiner}). In this framework, the
Black--Scholes option pricing method was first based on the geometric
Brownian motion \cite{BS,Merton}, and it was lately extended to include
white shot noise.

The Black-Sholes PDE (\ref{BS}) is usually derived from SDEs describing the
geometric Brownian motion (\ref{gbm}), with the stock-price solution given
by:
\begin{equation*}
s(t)= s(0)\mathrm{e}^{(\mu- \frac{1}{2} \sigma^2) t+ \sigma W(t)}.
\end{equation*}
In mathematical finance, derivation is usually performed using It\^{o} lemma
\cite{Ito} (assuming that the underlying asset obeys the It\^{o} SDE), while
in physics it is performed using Stratonovich interpretation (assuming that
the underlying asset obeys the Stratonovich SDE \cite{Stratonovich}) \cite%
{Gardiner,Perello}.

The PDE (\ref{BS}) resembles the backward Fokker--Planck equation (also
known as the Kolmogorov forward equation, in which the probabilities diffuse
outwards as time moves forwards) describes the time evolution of the
probability density function $p=p(t,x)$ for the position $x$ of a particle,
and can be generalized to other observables as well \cite{Kadanoff}. Its
first use was statistical description of Brownian motion of a particle in a
fluid. Applied to the option--pricing process $p=p(t,s)$ with \emph{drift} $%
D_{1}=D_{1}(t,s) $, \emph{diffusion} $D_{2}=D_{2}(t,s)$ and volatility $%
\sigma^2$, the forward Fokker--Planck equation reads:
\begin{equation*}
\partial _{t}p=\frac{1}{2}\partial _{ss}\left(
D_{2}\sigma^2p\right)-\partial _{s}\left( D_{1}p\right).
\end{equation*}
The corresponding backward Fokker--Planck equation (which is probabilistic
diffusion in reverse, i.e., starting at the final forecasts, the
probabilities diffuse outwards as time moves backwards) reads:
\begin{equation*}
\partial _{t}p=-\frac{1}{2}\sigma^2\partial _{ss}\left(
D_{2}p\right)-\partial _{s}\left( D_{1}p\right).
\end{equation*}

The solution of the PDE (\ref{BS}) depends on boundary conditions, subject
to a number of interpretations, some requiring minor transformations of the
basic BS equation or its solution.

\begin{figure}[htb]
\centering \includegraphics[width=7cm]{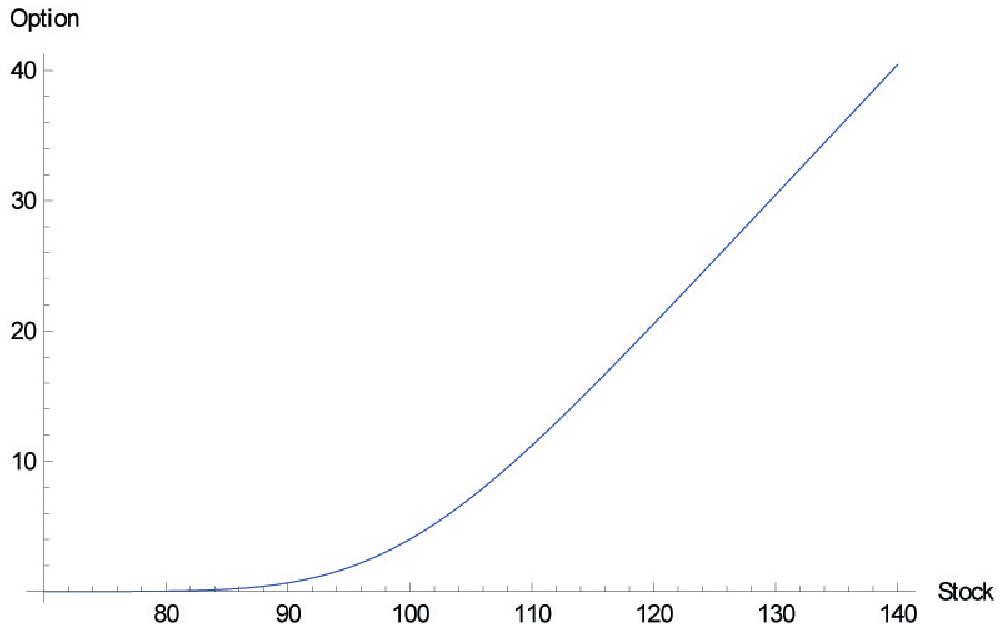} %
\includegraphics[width=7cm]{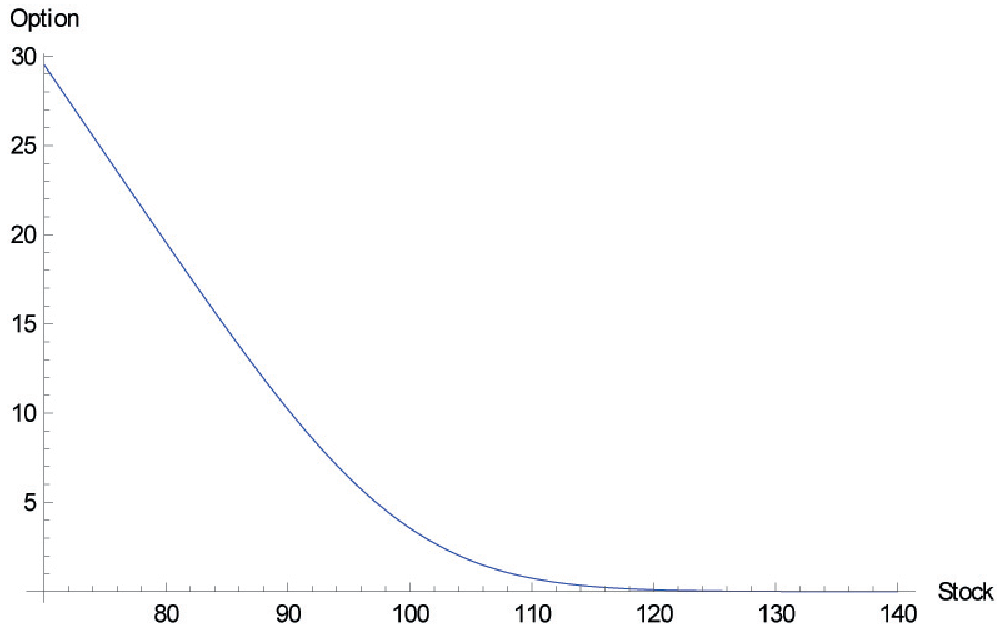}
\caption{European call (\protect\ref{call}) and put (\protect\ref{put})
options, as the solutions of the Black-Sholes PDE (\protect\ref{BS}). Used
parameters are: $\protect\sigma=0.3,\,r=0.05,\, k=100,\, \protect\delta=0.04$
\protect\cite{Kelly}.}
\label{callput}
\end{figure}

The basic equation (\ref{BS}) can be applied to a number of one-dimensional
models of interpretations of prices given to $u$, e.g., puts or calls, and
to $s$, e.g., stocks or futures, dividends, etc. The most important examples
are European call and put options (see Figure \ref{callput}), defined by:
\begin{eqnarray}
&&u_{\mathrm{Call}}(s,t) = s\,\mathcal{N}(\mathrm{d_1})\,\mathrm{e}%
^{-T\delta }-k\,\mathcal{N}(\mathrm{d_2})\,\mathrm{e}^{-rT},  \label{call} \\
&&u_{\mathrm{Put}}(s,t) = k\,\mathcal{N}(-\mathrm{d_2})\,\mathrm{e}^{-rT}-s\,%
\mathcal{N}(-\mathrm{d_1})\,\mathrm{e}^{-T\delta },  \label{put} \\
&&\,\mathcal{N}(\lambda ) = \frac{1}{2}\left( 1+\mathrm{erf}\left( \frac{%
\lambda }{\sqrt{2}}\right) \right) ,  \notag \\
&&\mathrm{d_1} = \frac{\ln \left( \frac{s}{k}\right) +T\left( r-\delta +%
\frac{\sigma ^{2}}{2}\right) }{\sigma \sqrt{T}}, \qquad \mathrm{d_2} = \frac{%
\ln \left( \frac{s}{k}\right) +T\left( r-\delta -\frac{\sigma ^{2}}{2}%
\right) }{\sigma \sqrt{T}},  \notag
\end{eqnarray}
where erf$(\lambda)$ is the (real-valued) error function, $k$ denotes the
strike price and $\delta$ represents the dividend yield. In addition, for
each of the call and put options, there are five Greeks (see, e.g. \cite%
{Kelly}), or sensitivities of the option-price with respect to the following
quantities:
\begin{enumerate}
\item The stock price -- Delta: $\Delta _{\mathrm{Call}}=\partial _{s}u_{%
\mathrm{Call}}\ $and $\Delta _{\mathrm{Put}}=\partial _{s}u_{\mathrm{Put}};$

\item The interest rate -- Rho: $\rho _{\mathrm{Call}}=\partial _{r}u_{%
\mathrm{Call}}\ $and $\rho _{\mathrm{Put}}=\partial _{r}u_{\mathrm{Put}};$

\item The volatility: Vega$_{\mathrm{Call}}=\partial _{\sigma }u_{\mathrm{%
Call}}\ $and Vega$_{\mathrm{Put}}=\partial _{\sigma }u_{\mathrm{Put}};$

\item The elapsed time since entering into the option -- Theta:\newline
$\Theta _{\mathrm{Call}}=\partial _{T}u_{\mathrm{Call}}\ $and $\Theta _{%
\mathrm{Put}}=\partial _{T}u_{\mathrm{Put}};$ ~and

\item The second partial derivative of the option-price with respect to the
stock price -- Gamma: $\Gamma _{\mathrm{Call}}=\partial _{ss}u_{\mathrm{Call}%
}\ $and $\Gamma _{\mathrm{Put}}=\partial _{ss}u_{\mathrm{Put}}$.
\end{enumerate}

In practice, the volatility $\sigma$ is the least known parameter in (\ref%
{BS}), and its estimation is generally the most important part of pricing
options. Usually, the volatility is given in a yearly basis, baselined to
some standard, e.g., 252 trading days per year, or 360 or 365 calendar days.
However, and especially after the 1987 crash, the geometric Brownian motion
model and the BS formula were unable to reproduce the option price data of
real markets.

The Black--Scholes model assumes that the underlying volatility is constant
over the life of the derivative, and unaffected by the changes in the price
level of the underlying. However, this model cannot explain long-observed
features of the implied volatility surface such as \emph{volatility smile}
and skew, which indicate that implied volatility does tend to vary with
respect to strike price and expiration. By assuming that the volatility of
the underlying price is a stochastic process itself, rather than a constant,
it becomes possible to model derivatives more accurately.

As an alternative, models of financial dynamics based on two-dimensional
diffusion processes, known as stochastic volatility (SV) models \cite{Fouque}, are being widely accepted as a reasonable explanation for many empirical
observations collected under the name of `stylized facts' \cite{Cont}. In
such models the volatility, that is, the standard deviation of returns,
originally thought to be a constant, is a random process coupled with the
return in a SDE of the form similar to (\ref{gbm}), so that they both form a
two-dimensional diffusion process governed by a pair of Langevin equations
\cite{Fouque,Perello08,Masoliver}.

Using the standard \emph{Kolmogorov probability} approach, instead of the
market value of an option given by the Black--Scholes equation (\ref{BS}),
we could consider the corresponding probability density function (PDF) given
by the backward Fokker--Planck equation (see \cite{Gardiner}).
Alternatively, we can obtain the same PDF (for the market value of a stock
option), using the \emph{quantum--probability} formalism \cite%
{ComplexDyn,QuLeap}, as a solution to a time--dependent linear \emph{%
Schr\"odinger equation} for the evolution of the complex--valued wave $\psi-$%
function for which the absolute square, $|\psi|^2,$ is the PDF (see \cite%
{Voit}).

In this paper, I will go a step further and propose a novel general
quantum--probability based,\footnote{%
Note that the domain of validity of the `quantum probability' is not
restricted to the microscopic world \cite{Ume93}. There are macroscopic
features of classically behaving systems, which cannot be explained without
recourse to the quantum dynamics (see \cite{FreVit06} and references
therein).} option--pricing model, which is both \emph{nonlinear} (see \cite%
{Trippi,Rothman,Ammann,HighDyn}) and \emph{adaptive} (see \cite%
{Tse,Ingber,GeoDyn,ComNonlin}). More precisely, I propose a \emph{quantum
neural computation} \cite{QnnBk} approach to option price modelling, based
on the nonlinear Schr\"{o}dinger (NLS) equation with adaptive parameters.

\section{Adaptive nonlinear Schr\"{o}dinger equation model}

This new adaptive wave--form approach to financial modelling is motivated by:
\begin{enumerate}
\item Modern adaptive markets hypothesis of A. Lo \cite{Lo1,Lo2};

\item My adaptive path integral approach to human cognition \cite%
{IA,NeuQuant,IAY};

\item Elliott wave (fractal) market theory \cite%
{Elliott1,Elliott2,MandelSciAm}; ~and

\item My recent monograph: `Quantum Neural Computation' \cite{QnnBk}, as
well as papers on entropic crowd modelling based on the concept of \emph{%
controlled Brownian motion} \cite{IvReid1,IvReid2,IvReid3}.
\end{enumerate}

To satisfy both efficient and behavioral markets, as well as their essential
nonlinear complexity, I propose an adaptive, wave--form, nonlinear and
stochastic option--pricing model with stock price $s,$ volatility $\sigma $
and interest rate $r$. The model is formally defined as a complex-valued,
focusing (1+1)--NLS equation, defining the \emph{option--price wave function}
$\psi =\psi (s,t)$, whose absolute square\ $|\psi (s,t)|^{2}$ represents the
probability density function (PDF) for the option price in terms of the
stock price and time. In natural quantum units, this NLS equation reads:
\begin{equation}
\mathrm{i}\partial _{t}\psi =-\frac{1}{2}\sigma \partial _{ss}\psi -\beta
|\psi |^{2}\psi ,\qquad (\mathrm{i}=\sqrt{-1})  \label{nlsGen}
\end{equation}%
where dispersion frequency coefficient $\sigma $ is the volatility (which
can be either constant or stochastic process itself), while Landau
coefficient $\beta =\beta (r,w)$ represents the adaptive market potential, which
is, in the simplest nonadaptive scenario, equal to the interest rate $r$,
while in the adaptive case depends on the set of adjustable parameters $\{w_{i}\}$.
In this case, $\beta (r,w)$ can be related to the market \emph{temperature}
(which obeys Boltzman distribution \cite{KleinertBk}). The term
$V(\psi )=-\beta |\psi |^{2}$ represents the $\psi -$dependent potential
field. Physically, the NLS equation (\ref{nlsGen}) describes a nonlinear
wave--packet defined by the complex-valued wave function $\psi (s,t)$ of
real space and time parameters. In the present context, the space-like
variable $s$ denotes the stock (asset) price.

\subsection{Analytical NLS--solution}

NLS equation can be exactly solved using the power series expansion method
\cite{LiuEtAl01,LiuFan05} of \emph{Jacobi elliptic functions} \cite{AbrSte}.
\begin{figure}[htb]
\centering \includegraphics[width=10cm]{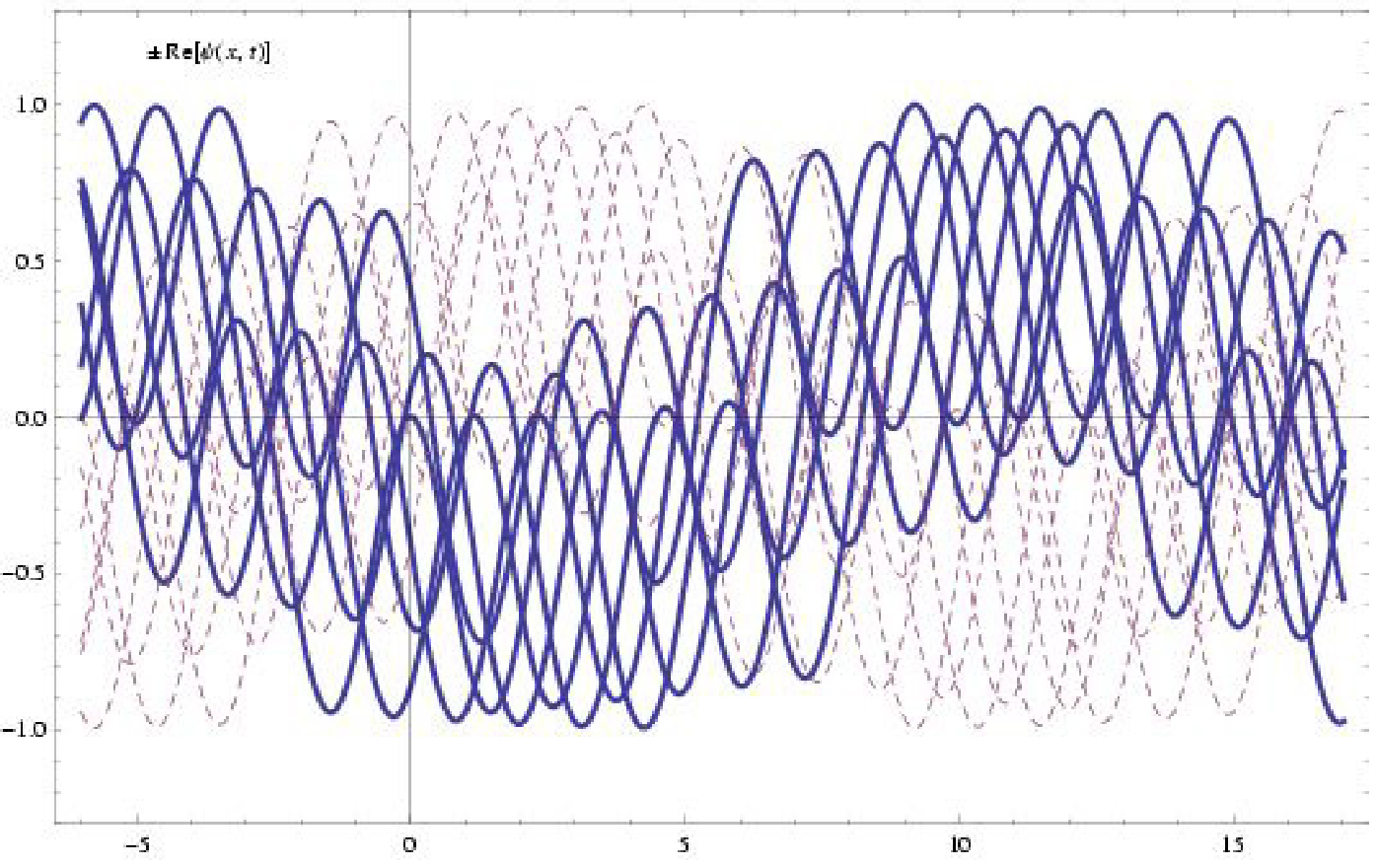}
\caption{The general Jacobi sine solution (\protect\ref{sn1}) of the NLS
equation (\protect\ref{nlsGen}) with $k=1.2,\,m=0.5$, $\protect\sigma =%
\protect\beta =1$, for $t\in (0,5)$ and $s\in (-7,18)$. Thick line
represents +Re$[\protect\psi (s,t)]$, while dash line represents --Re$[%
\protect\psi (s,t)]$.}
\label{Sn1}
\end{figure}
\begin{figure}[htb]
\centering \includegraphics[width=10cm]{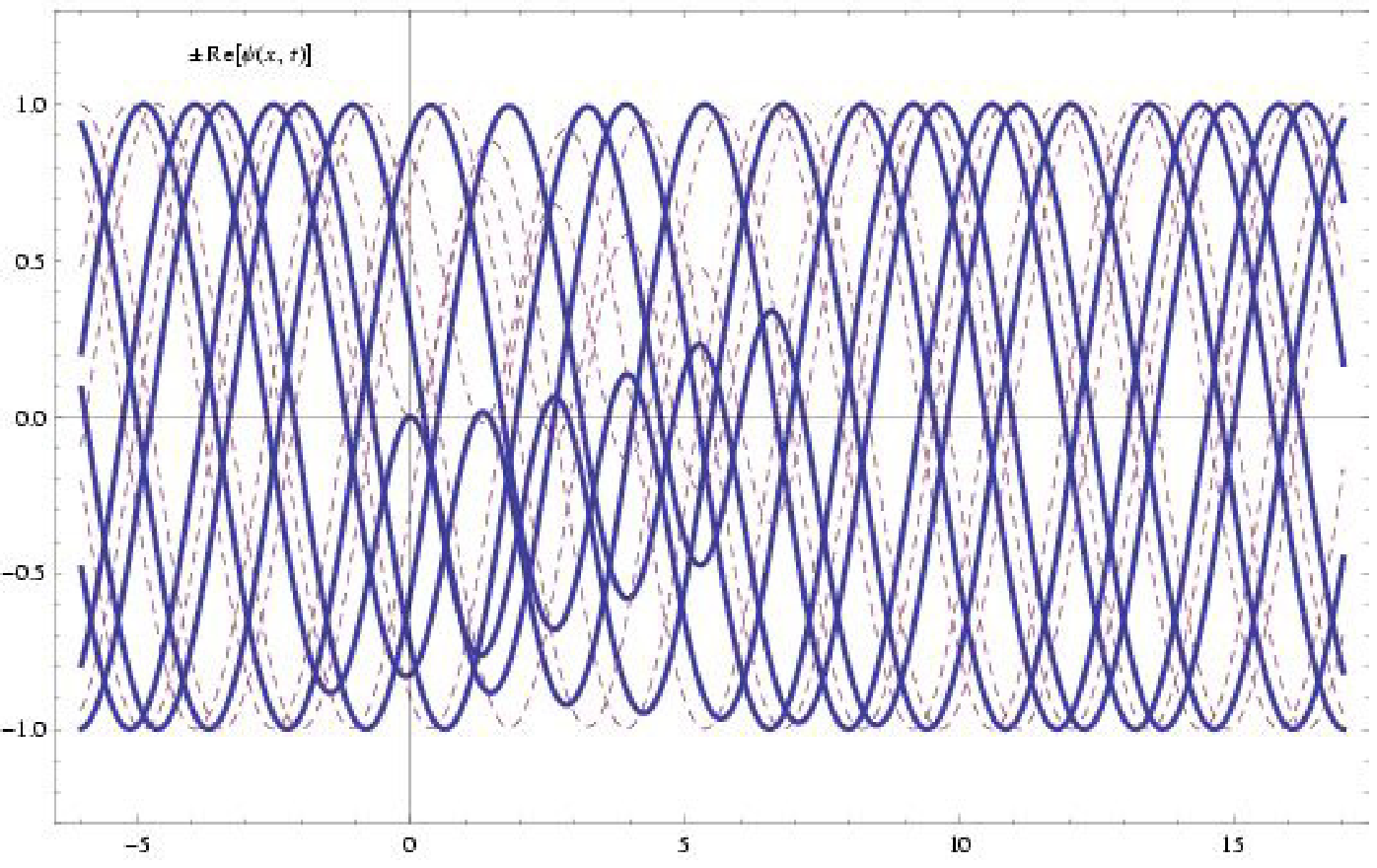}
\caption{The dark shock-wave solution (\protect\ref{tanh1}) of the NLS
equation (\protect\ref{nlsGen}) with $k=1.2$, $\protect\sigma =\protect\beta %
=1$, for $t\in (0,5)$ and $s\in (-7,18)$. Thick line represents +Re$[\protect%
\psi (s,t)]$, while dash line represents --Re$[\protect\psi (s,t)]$.}
\label{Tanh1}
\end{figure}
\begin{figure}[htb]
\centering \includegraphics[width=10cm]{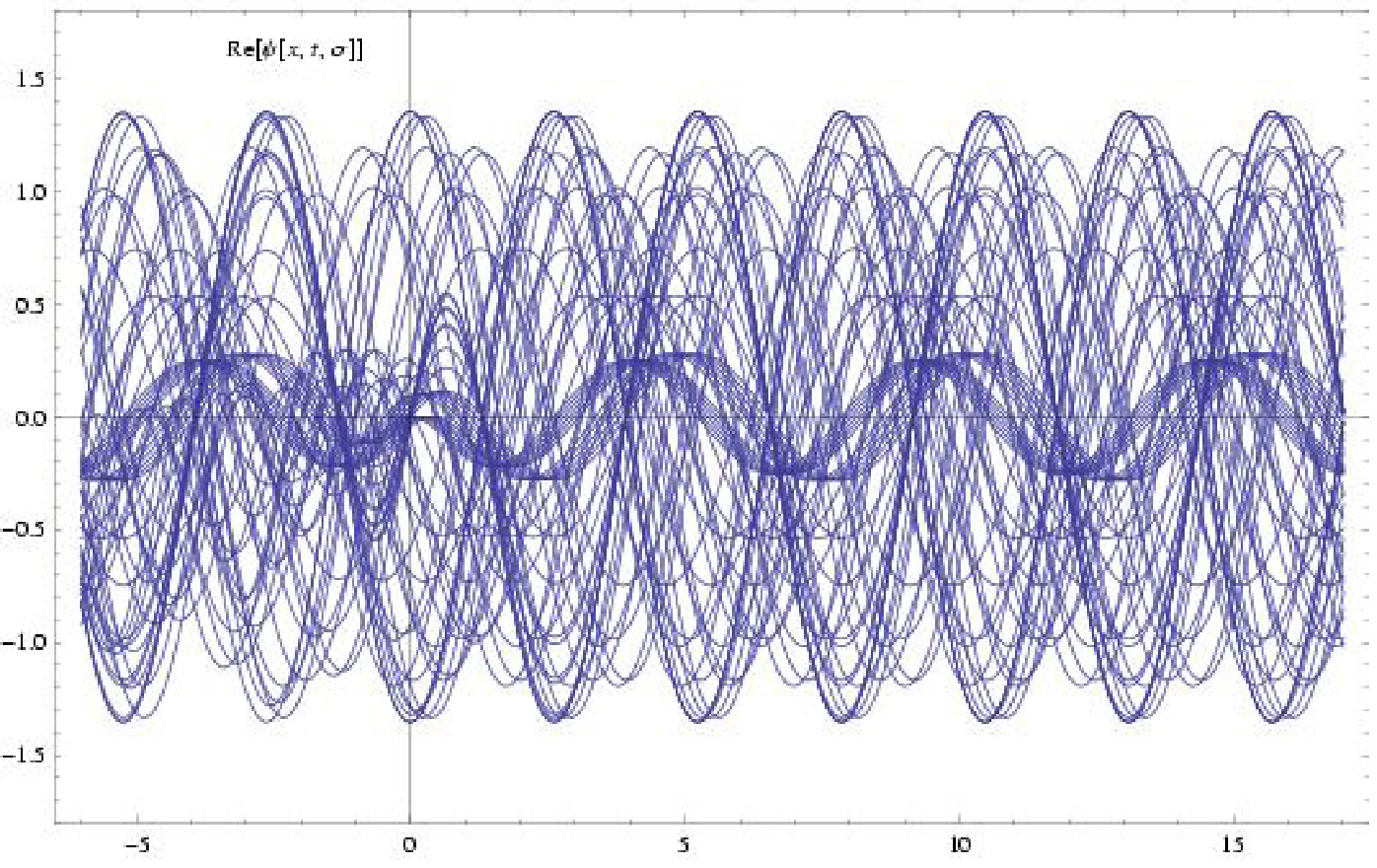}
\caption{The +tanh solution from Figure \protect\ref{Tanh1}, with stochastic
volatility $\protect\sigma _{t}$ (random walk).}
\label{TanhVolat1}
\end{figure}

In case of $\beta \ll 1$, $V(\psi )\rightarrow 0$, so equation (\ref{nlsGen}%
) can be approximated by a linear wave packet, defined by a continuous
superposition of de Broglie's plane waves, associated with a free quantum
particle. This linear wave packet is defined by the simple linear Schr\"{o}%
dinger equation with zero potential energy (in natural units):
\begin{equation}
\mathrm{i}\partial _{t}\psi =-\frac{1}{2}\partial _{ss}\psi .  \label{sch1}
\end{equation}%
Thus, we consider the $\psi -$function describing a single de Broglie's
plane wave, with the wave number (or, momentum) $k$ and circular frequency $%
\omega $:
\begin{equation}
\psi (s,t)=\phi (\xi )\,\mathrm{e}^{\mathrm{i}(ks-\omega t)},\qquad \text{%
with \ }\xi =s-\sigma kt\text{ \ and \ }\phi (\xi )\in \mathbb{R}.
\label{subGen}
\end{equation}%
Its substitution into the linear Schr\"{o}dinger equation (\ref{sch1}) gives
the linear harmonic oscillator ODE, whose eigenvalues are natural
frequencies of (\ref{sch1}) and the solution is given by a Fourier sine or
cosine series (see, e.g. \cite{Griffiths,Thaller}).
\begin{figure}[htb]
\centering \includegraphics[width=10cm]{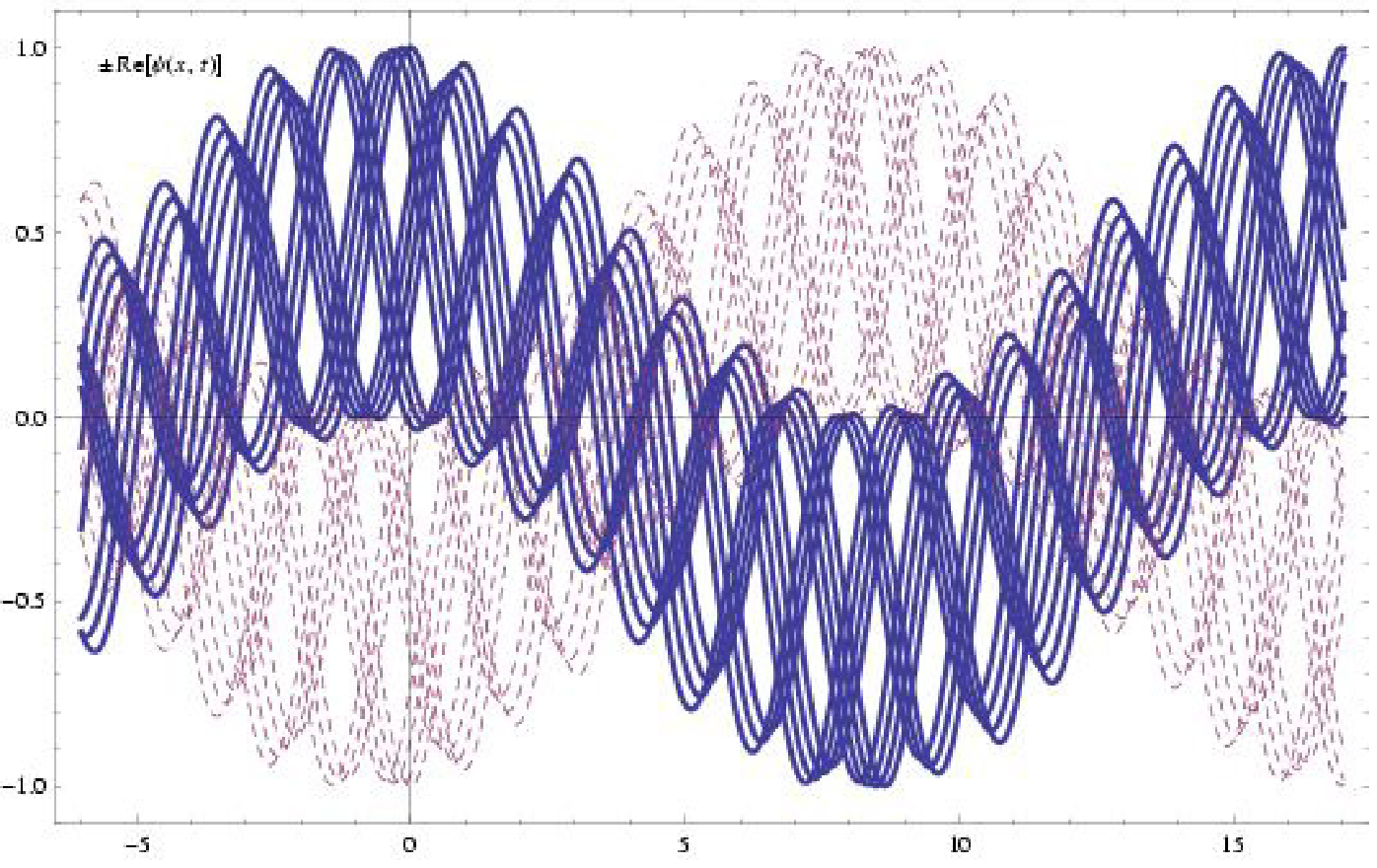}
\caption{The general Jacobi cosine solution (\protect\ref{cn1}) of the NLS
equation (\protect\ref{nlsGen}) with $k=1.2,\,m=0.5$, $\protect\sigma =%
\protect\beta =1$, for $t\in (0,10)$ and $s\in (-7,18)$. Thick line
represents +Re$[\protect\psi (s,t)]$, while dash line represents --Re$[%
\protect\psi (s,t)]$.}
\label{Cn1}
\end{figure}
\begin{figure}[htb]
\centering \includegraphics[width=10cm]{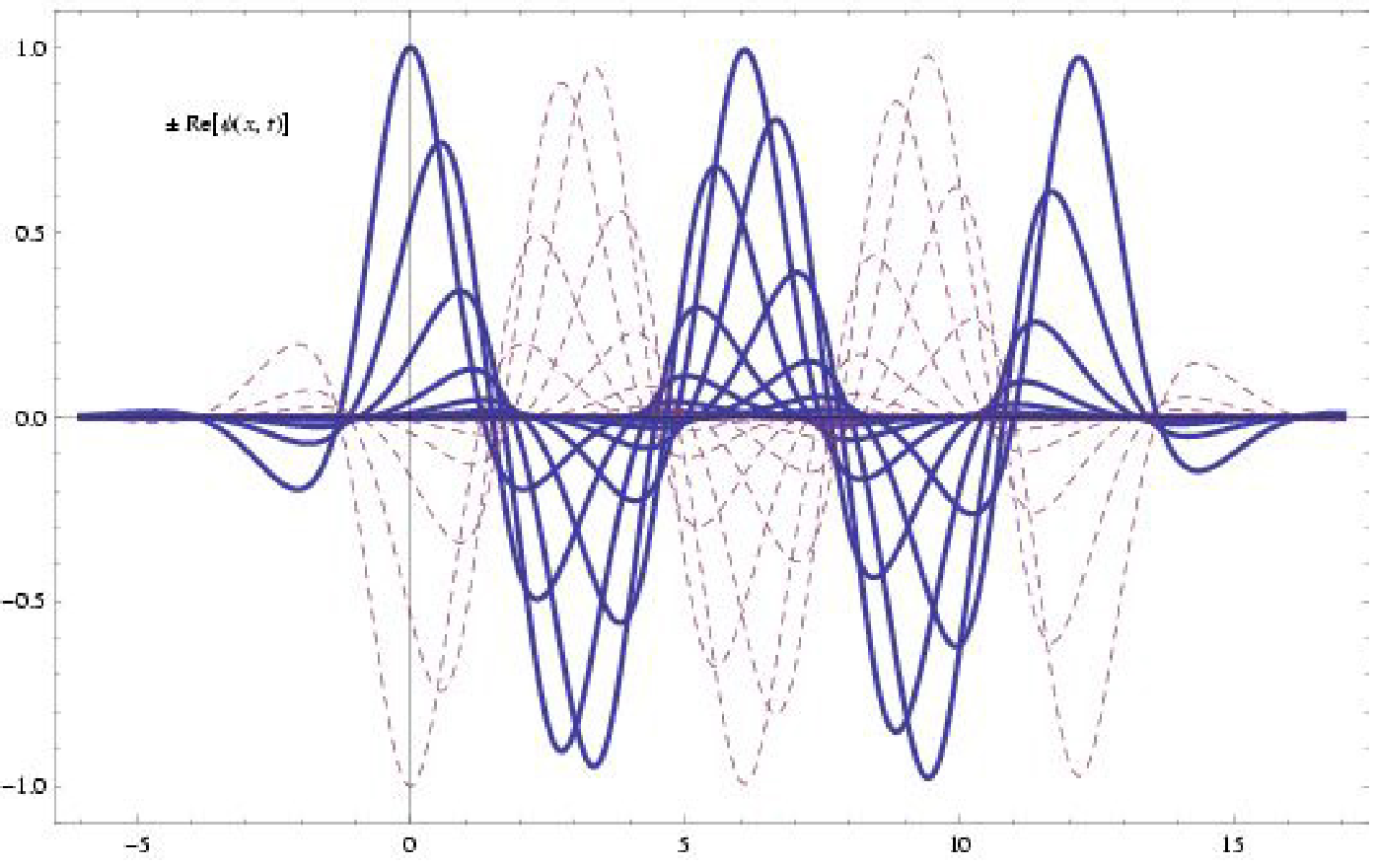}
\caption{The bright solitary-wave solution (\protect\ref{sech1}) of the NLS
equation (\protect\ref{nlsGen}) with $k=1.2$, $\protect\sigma =\protect\beta %
=1$, for $t\in (0,10)$ and $s\in (-7,18)$. Thick line represents +Re$[%
\protect\psi (s,t)]$, while dash line represents --Re$[\protect\psi (s,t)]$.}
\label{Sech1}
\end{figure}
\begin{figure}[htb]
\centering \includegraphics[width=10cm]{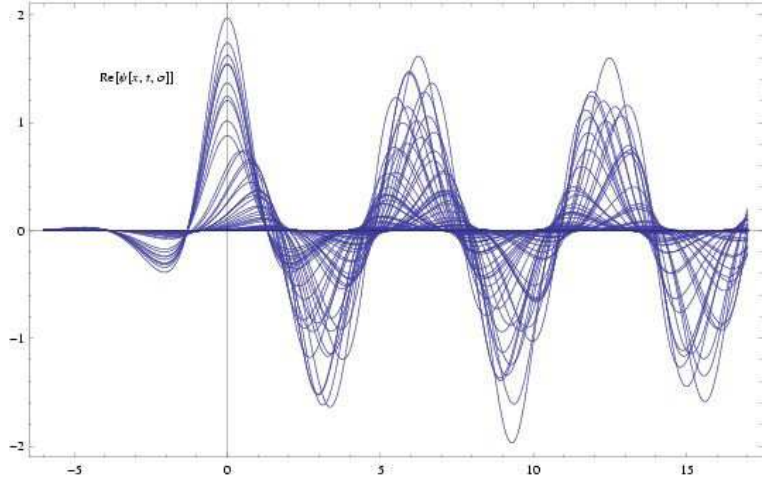}
\caption{The +sech soliton from Figure \protect\ref{Sech1}, with stochastic
volatility $\protect\sigma _{t}$ (random walk).}
\label{SechVolat1}
\end{figure}

Similarly, substituting (\ref{subGen}) into the NLS equation (\ref{nlsGen}),
we obtain a nonlinear oscillator ODE:
\begin{equation}
\phi ^{\prime \prime }(\xi )+[\omega -\frac{1}{2}\sigma k^{2}]\,\phi (\xi
)+\beta \phi ^{3}(\xi )=0.  \label{15e}
\end{equation}%
Following \cite{LiuFan05}, I suppose that a solution $\phi (\xi )$ for (\ref%
{15e}) can be obtained as a linear expansion
\begin{equation}
\phi (\xi )=a_{0}+a_{1}\mathrm{sn}(\xi ),  \label{17e}
\end{equation}%
where $\mathrm{sn}(s)=\mathrm{sn}(s,m)$ are Jacobi elliptic sine functions
with \textit{elliptic modulus} $m\in \lbrack 0,1]$, such that $\mathrm{sn}%
(s,0)=\sin (s)\ $and $\mathrm{sn}(s,1)=\mathrm{\tanh }(s)$.\footnote{%
For example, the general pendulum equation:
\begin{equation*}
\alpha ^{\prime \prime }(t,\phi )+\sin [\alpha (t,\phi )]=0
\end{equation*}%
has the elliptic solution:
\begin{equation*}
\alpha (t,\phi )=2\sin ^{-1}\left[ \sin \left( \frac{\phi }{2}\right) \right]
\,\mathrm{sn}\left[ t,\sin ^{2}\left( \frac{\phi }{2}\right) \right] .
\end{equation*}%
} Using standard identities with associated elliptic cosine functions $%
\mathrm{cn}(\xi )$ and elliptic functions of the third kind $\mathrm{dn}(\xi
)$, we have
\begin{eqnarray}
\phi ^{\prime }(\xi ) &=&a_{1}\mathrm{cn}(\xi )\,\mathrm{dn}(\xi ),  \notag
\\
\phi ^{\prime \prime }(\xi ) &=&-a1\{\mathrm{sn}(\xi )[1-m^{2}\mathrm{sn}%
^{2}(\xi )]+m^{2}\mathrm{sn}(\xi )[1-\mathrm{sn}^{2}(\xi )]\}.  \label{19e}
\end{eqnarray}%
Substituting (\ref{17e}) and (\ref{19e}) into (\ref{15e}), after doing some
algebra, we get
\begin{equation}
a_{0}=0,\qquad a_{1}=\pm m\sqrt{\frac{-\sigma }{\beta }},\qquad \omega =%
\frac{1}{2}(1+m^{2}+k^{2}),  \label{21e}
\end{equation}%
which, substituted into the nonlinear oscillator (\ref{15e}), gives
\begin{eqnarray*}
\phi (\xi ) &=&\pm m\sqrt{\frac{-\sigma }{\beta }}\,\mathrm{sn}(\xi ),\qquad
~\text{for~~}m\in \lbrack 0,1];~~\text{and} \\
\phi (\xi ) &=&\pm \sqrt{\frac{-\sigma }{\beta }}\,\mathrm{\tanh }(\xi
),\qquad \text{for~~}m=1.
\end{eqnarray*}%
Using the substitutions (\ref{subGen}) and (\ref{21e}), we now obtain the
exact periodic solution of (\ref{nlsGen}) as
\begin{eqnarray}
\psi _{1}(s,t) &=&\pm m\sqrt{\frac{-\sigma }{\beta (w)}}\,\mathrm{sn}%
(s-\sigma kt)\,\mathrm{e}^{\mathrm{i}[ks-\frac{1}{2}\sigma
t(1+m^{2}+k^{2})]},\qquad ~~\text{for~~}m\in \lbrack 0,1);  \label{sn1} \\
\psi _{2}(s,t) &=&\pm \sqrt{\frac{-\sigma }{\beta (w)}}\,\mathrm{\tanh }%
(s-\sigma kt)\,\mathrm{e}^{\mathrm{i}[ks-\frac{1}{2}\sigma
t(2+k^{2})]},\qquad \qquad \text{for~~}m=1,  \label{tanh1}
\end{eqnarray}%
where (\ref{sn1}) defines the general solution (see Figure \ref{Sn1}), while
(\ref{tanh1}) defines the \emph{envelope shock-wave}\footnote{%
A shock wave is a type of fast-propagating nonlinear disturbance that
carries energy and can propagate through a medium (or, field). It is
characterized by an abrupt, nearly discontinuous change in the
characteristics of the medium. The energy of a shock wave dissipates
relatively quickly with distance and its entropy increases. On the other
hand, a soliton is a self-reinforcing nonlinear solitary wave packet that
maintains its shape while it travels at constant speed. It is caused by a
cancelation of nonlinear and dispersive effects in the medium (or, field).}
(or, `dark soliton') solution (Figure \ref{Tanh1}) of the NLS equation (\ref%
{nlsGen}). The same shock-wave solution with stochastic volatility $\sigma
_{t}$ (defined as a simple random walk) is given in Figure \ref{TanhVolat1}.

Alternatively, if we seek a solution $\phi (\xi )$ as a linear expansion of
Jacobi elliptic cosine functions, such that $\mathrm{cn}(s,0)=\cos (s)$ and $%
\mathrm{cn}(s,1)=\mathrm{sech}(s)$,\footnote{%
A closely related solution of an anharmonic oscillator ODE:
\begin{equation*}
\phi ^{\prime \prime }(s)+\phi (s)+\phi ^{3}(s)=0
\end{equation*}%
is given by
\begin{equation*}
\phi (s)=\sqrt{\frac{2m}{1-2m}}\,\text{cn}\left( \sqrt{1+\frac{2m}{1-2m}}%
~s,\,m\right) .
\end{equation*}%
} in a linear form:
\begin{equation*}
\phi (\xi )=a_{0}+a_{1}\mathrm{cn}(\xi ),
\end{equation*}%
then we get
\begin{eqnarray}
\psi _{3}(s,t) &=&\pm m\sqrt{\frac{\sigma }{\beta (w)}}\,\mathrm{cn}%
(s-\sigma kt)\,\mathrm{e}^{\mathrm{i}[ks-\frac{1}{2}\sigma
t(1-2m^{2}+k^{2})]},\qquad \text{for~~}m\in \lbrack 0,1);  \label{cn1} \\
\psi _{4}(s,t) &=&\pm \sqrt{\frac{\sigma }{\beta (w)}}\,\mathrm{sech}%
(s-\sigma kt)\,\mathrm{e}^{\mathrm{i}[ks-\frac{1}{2}\sigma
t(k^{2}-1)]},\qquad \qquad \text{for~~}m=1,  \label{sech1}
\end{eqnarray}%
where (\ref{cn1}) defines the general solution (Figure \ref{Cn1}), while (%
\ref{sech1}) defines the \emph{envelope solitary-wave} (or, `bright
soliton') solution (Figure \ref{Sech1}) of the NLS equation (\ref{nlsGen}).
The same soliton solution with stochastic volatility $\sigma_t$ (a simple
random walk) is given in Figure \ref{TanhVolat1}.

In all four solution expressions (\ref{sn1}), (\ref{tanh1}), (\ref{cn1}) and
(\ref{sech1}), the adaptive potential $\beta (w)$ is yet to be calculated
using either unsupervised Hebbian learning, or supervised
Levenberg--Marquardt algorithm (see, e.g. \cite{NeuFuz,CompMind}). In this
way, the NLS equation (\ref{nlsGen}) resembles the `quantum
stochastic-filtering neural network' model of \cite%
{BeheraFilt,Behera05,Behera06}. While the authors of the prior quantum
neural network performed only numerical (finite-difference) simulations of
their model, this paper provides theoretical foundation (of both single NLS
network and coupled NLS network) with closed-form analytical solutions. Any
kind of numerical analysis can be easily performed using above closed-form
solutions $\psi _{i}(s,t),~~(i=1,...,4)$.

\subsection{Fitting the Black--Scholes model using adaptive NLS--PDF}

The adaptive NLS--PDFs of the shock-wave type (\ref{tanh1}) can be used
to fit the Black--Scholes call and put options. Specifically, I have used
the spatial part of (\ref{tanh1}),
\begin{equation}
\phi (s)=\left\vert \sqrt{\frac{\sigma }{\beta }}\tanh (s-kt\sigma
)\right\vert {}^{2}, \label{kink2}
\end{equation}
where the adaptive
market--heat potential $\beta (r,w)$ is chosen as:
\begin{equation}
\beta (r,w)=r\sum_{i=1}^{n}w_{1}^{i}\,\text{erf}\left( \frac{w_{2}^{i}s}{%
w_{3}^{i}}\right)  \label{betaW}
\end{equation}%
\begin{figure}[htb]
\centering \includegraphics[width=10cm]{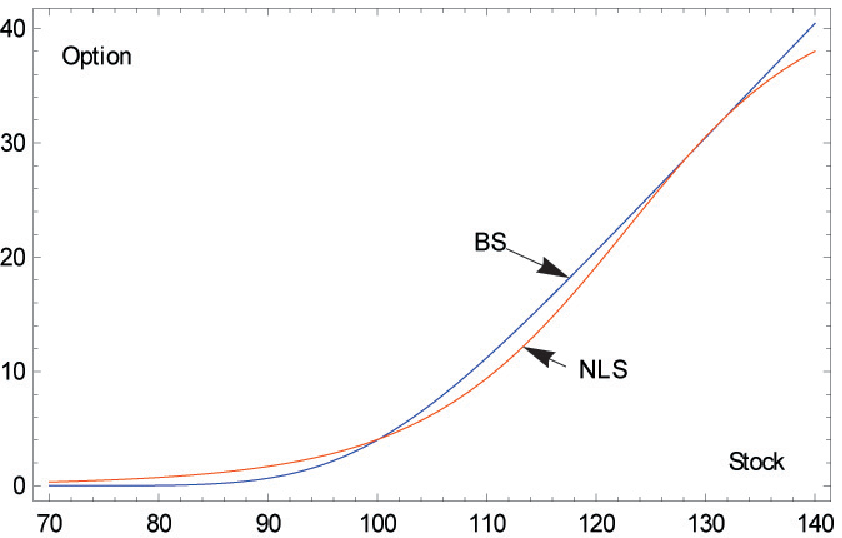}
\caption{Fitting the Black--Scholes call option with $\protect\beta (w)$%
-adaptive PDF of the shock-wave NLS-solution (\protect\ref{tanh1}).}
\label{fitCall}
\end{figure}
\begin{figure}[htb]
\centering \includegraphics[width=10cm]{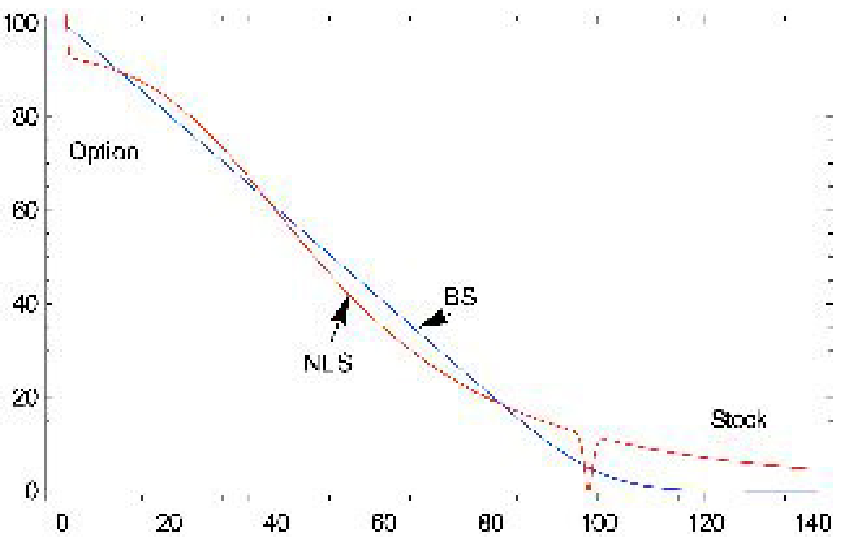}
\caption{Fitting the Black--Scholes put option with $\protect\beta (w)$
--adaptive PDF of the shock-wave NLS $\protect\psi _{2}(s,t)$ solution (%
\protect\ref{tanh1}). Notice the kink near $s=100$.}
\label{fitPut}
\end{figure}

The following parameter estimates where obtained using 100 iterations of the
Levenberg--Marquardt algorithm. In case of the call option fit (see Figure %
\ref{fitCall}), $n=5$,

$w_{1}^{1}=24.8952,~ w_{2}^{1}=-78112.3,~ w_{3}^{1}=-48178.3,$

$w_{1}^{2}=24.8951,~ w_{2}^{2}=-78112.3,~ w_{3}^{2}=-48178.3,$

$w_{1}^{3}=24.895,~ w_{2}^{3}=-78112.3,~ w_{3}^{3}=-48178.3,$

$w_{1}^{4}=-37.3927,~ w_{2}^{4}=-3108.08,~ w_{3}^{4}=-1520633,$

$w_{1}^{5}=-37.2757,~ w_{2}^{5}=-3968.35,~ w_{3}^{5}=-159782.$

$\sigma^{\mathrm{call}}_{\mathrm{NLS}}=-0.119341\sigma_{\mathrm{BS}}$, $k^{%
\mathrm{call}}_{\mathrm{NLS}}=0.0156422k_{\mathrm{BS}}$, $T^{\mathrm{call}}_{%
\mathrm{NLS}}=15.6423T_{\mathrm{BS}}.$\newline

In case of the put option fit (see Figure (\ref{fitPut})), $n=3$,

$w_a^1= 0.000222367,~w_b^1=82032.8,~w_c^1= 63876.9,$

$w_a^2=-0.428113,~w_b^2= 439.148,~w_c^2=205780.0,$

$w_a^3= 4.70615,~w_b^3=27.1558,~w_c^3= 139805.0$

$\sigma^{\mathrm{put}}_{\mathrm{NLS}}=-0.003444\sigma_{\mathrm{BS}}$, $k^{\mathrm{put}}_{\mathrm{NLS}}=-3.10354k_{\mathrm{BS}}$, $T^{\mathrm{put}}_{\mathrm{NLS}}=-3103.54T_{\mathrm{BS}}.$\\

\noindent As can be seen from Figure (\ref{fitPut}) there is a kink near $s=100$. This kink, which is a natural characteristic of the spatial shock-wave (\ref{kink2}), can be smoothed out by taking the sum of the spatial parts of the shock-wave NLS-solution (\ref{tanh1}) and the soliton NLS-solution (\ref{sech1}) as:
\begin{equation}
\phi (s)=\left\vert \sqrt{\frac{\sigma }{\beta }}\left[d_{1}\tanh
(s-kt\sigma )+d_{2}\,\text{sech}(s-kt\sigma )\right] \right\vert {}^{2}.
\label{kinksech}
\end{equation}
In this case, using 100 iterations of the
Levenberg--Marquardt algorithm, the following parameter estimates where obtained:\\

$w_1^1=-0.00190885,\,w_2^1= 6798.78,\,w_3^1= 5329.46,\,$

$w_1^2= 18.1757,\,w_2^2=23.5253,\,w_3^2= 18354.9,\,$

$w_1^3= -71.7315,\,w_2^3= 4.15999,\,w_3^3=12807.2,\,$

$d_1= 0.345078,\,d_2= -12.3948.$

$\sigma^{\mathrm{put}}_{\mathrm{NLS}} = -0.247932_{\mathrm{BS}},\,k^{\mathrm{put}}_{\mathrm{NLS}}= 0.260764k_{\mathrm{BS}},\,T^{\mathrm{put}}_{\mathrm{NLS}}= 260.764T_{\mathrm{BS}}.$
\begin{figure}[htb]
\centering \includegraphics[width=10cm]{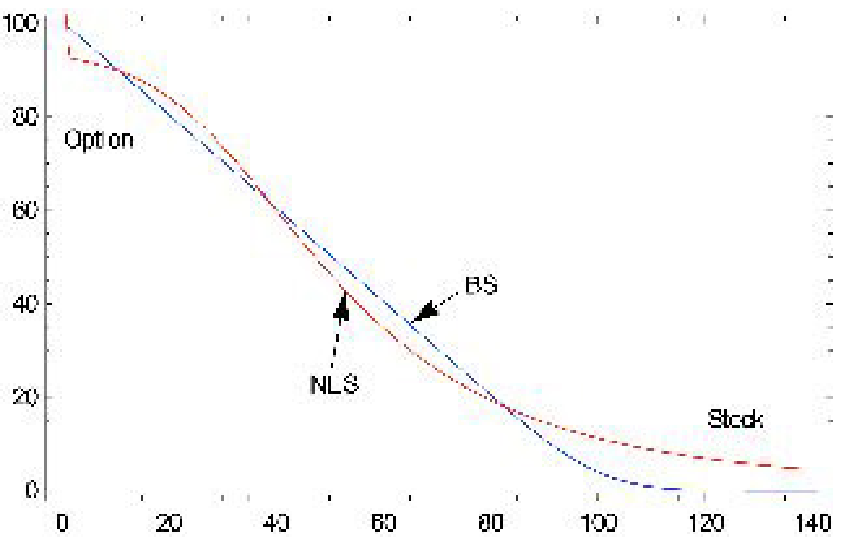}
\caption{Smoothing out the kink in the put option fit, by combining the shock-wave solution with the soliton solution, as defined by (\ref{kinksech}).}
\label{tanSechPut}
\end{figure}

The adaptive NLS--based Greeks can now be defined, using $\beta =r$ and
above modified $(\sigma ,\,k,\,t)$ values, by the following partial
derivatives of the spatial part of the shock-wave solution (\ref{tanh1}):%
\newline

\noindent$\mathrm{Delta}=\partial_s\phi(s)=2\sqrt{-\frac{\sigma }{r}}\sqrt{%
\left| \frac{\sigma }{r}\right| }\text{sech}^{2}(s-kt\sigma )|\tanh
(s-kt\sigma )|\text{abs}^{\prime}\left( \sqrt{-\frac{\sigma }{r}}\tanh
(s-kt\sigma )\right) ,$\newline

\noindent$\mathrm{Gamma}=\partial_ss\phi(s)=-\frac{2\text{sech}%
^{4}(s-kt\sigma )}{r}[\sigma $abs$^{\prime }\left( \sqrt{-\frac{\sigma }{r}}%
\tanh (s-kt\sigma )\right) ^{2}$

$+~\sqrt{\left| \frac{\sigma }{r}\right| }|\tanh (s-kt\sigma )|\{\sigma $abs$%
^{\prime \prime }\left( \sqrt{-\frac{\sigma }{r}}\tanh (s-kt\sigma )\right)$

$+~r\sqrt{-\frac{\sigma }{r}}\sinh (2s-2kt\sigma )$abs$^{\prime }\left(
\sqrt{-\frac{\sigma }{r}}\tanh (s-kt\sigma )\right) \}],$\newline

\noindent$\mathrm{Vega}=\partial_\sigma\phi(s)=\frac{\sqrt{-\frac{\sigma }{r}%
}\sqrt{\left| \frac{\sigma }{r}\right| }|\tanh (s-kt\sigma )|\left( \tanh
(s-kt\sigma )-2kt\sigma \text{sech}^{2}(s-kt\sigma )\right) \text{abs}%
^{\prime }\left( \sqrt{-\frac{\sigma }{r}}\tanh (s-kt\sigma )\right) }{%
\sigma },$\newline

\noindent$\mathrm{Rho}=\partial_r\phi(s)=\frac{\left( -\frac{\sigma }{r}%
\right) ^{3/2}\sqrt{\left| \frac{\sigma }{r}\right| }\tanh (s-kt\sigma
)|\tanh (s-kt\sigma )|\text{abs}^{\prime }\left( \sqrt{-\frac{\sigma }{r}}%
\tanh (s-kt\sigma )\right) }{\sigma },$\newline

\noindent$\mathrm{Theta}=\partial_t\phi(s)=2 k r \left(-\frac{\sigma }{r}%
\right)^{3/2} \sqrt{\left|\frac{\sigma }{r}\right|} \text{sech}^2(s-k t
\sigma ) |\tanh (s-k t \sigma )| \text{abs}^{\prime}\left(\sqrt{-\frac{%
\sigma }{r}} \tanh (s-k t \sigma )\right),$\newline

\noindent where $\text{abs}^{\prime }(z)$ denotes the partial derivative of
the absolute value upon the corresponding variable $z$.

\subsection{Coupled adaptive NLS--system for volatility + option-price
evolution modelling}

For the purpose of including a \emph{controlled stochastic volatility}\footnote{Controlled stochastic volatility here represents volatility evolving in a stochastic manner but within the controlled boundaries.} into the adaptive--wave model (\ref{nlsGen}), the full bidirectional quantum neural computation model for option price forecasting can be represented as a self-organized system of two coupled
self-focusing NLS equations: one defining the \emph{option--price wave
function} $\psi =\psi (s,t)$ and the other defining the \emph{volatility
wave function} $\sigma =\sigma (s,t)$. The two NLS equations are
coupled so that the volatility $\sigma$ is a parameter in the option--price
NLS, while the option--price $\psi$ is a parameter in the volatility NLS. In
addition, both processes evolve in a common self--organizing \emph{market
heat potential}, so they effectively represent an \emph{adaptively controlled Brownian behavior} of a hypothetical financial market.

Formally, I here propose an adaptive, symmetrically coupled, volatility +
option--pricing model (with interest rate $r$ and Hebbian learning rate $c$), which represents a bidirectional spatio-temporal associative
memory. The model is defined by the following coupled--NLS+Hebb
system:
\begin{eqnarray}
\text{Volatility NLS :}\quad \mathrm{i}\partial _{t}\sigma &=&-\frac{1}{2}%
\partial _{ss}\mathcal{\sigma }-\beta \left( |\mathcal{\sigma }|^{2}+|\psi
|^{2}\right) \mathcal{\sigma },  \label{stochVol} \\
\text{Option price NLS :}\quad \mathrm{i}\partial _{t}\psi &=&-\frac{1}{2}%
\partial _{ss}\psi -\beta \left( |\mathcal{\sigma }|^{2}+|\psi |^{2}\right)
\psi ,  \label{stochPrice} \\
\text{with :}~~\beta (r,w) &=&r\sum_{i=1}^{N}w_{i}g_{i},\qquad \text{and}
\notag \\
\text{Adaptation ODE :}\quad \dot{w}_{i} &=&-w_{i}+c|\sigma |g_{i}|\psi |.
\label{Hebb1}
\end{eqnarray}%
In this coupled model, the $\sigma $--NLS (\ref{stochVol}) governs the
$(s,t)-$evolution of stochastic volatility, which plays the role of a nonlinear
coefficient in (\ref{stochPrice}); the $\psi $--NLS (\ref%
{stochPrice}) defines the $(s,t)-$evolution of option price, which plays the role of
a nonlinear coefficient in (\ref{stochVol}). The purpose of this coupling is
to generate a \emph{leverage effect}, i.e. stock volatility is (negatively)
correlated to stock returns\footnote{%
The hypothesis that financial leverage can explain the leverage effect was
first discussed by F. Black \cite{Bl76}.} (see, e.g. \cite{Roman}). The $w-$%
ODE (\ref{Hebb1}) defines the $(\sigma ,\psi )-$coupling based continuous
Hebbian learning with the learning rate $c.$ The adaptive market--heat potential $\beta (r,w)$, previously defined by (\ref{betaW}), is now generalized into a scalar product of the `synaptic weight' vector $w_{i}$ and the Gaussian kernel vector $g_{i}$, yet to be
defined.

The bidirectional associative memory model (\ref{stochVol})--(\ref{stochPrice})--(\ref{Hebb1}) effectively performs quantum neural computation \cite{QnnBk}, by giving a spatio-temporal and quantum generalization of Kosko's
BAM family of neural networks \cite{Kosko1,Kosko2}. In addition, the
shock-wave and solitary-wave nature of the coupled NLS equations may
describe brain-like effects frequently occurring in financial markets:
volatility/price propagation, reflection and collision of shock and solitary
waves (see \cite{Han}).

The coupled NLS-system (\ref{stochVol})--(\ref{stochPrice}), without an
embedded $w-$learning (i.e., for constant $\beta=r$ -- the interest rate),
actually defines the well-known \emph{Manakov system}, which was proven by
S. Manakov in 1973 \cite{manak74} to be completely integrable, by the
existence of infinite number of involutive integrals of motion. It admits
`bright' and `dark' soliton solutions. Manakov system has been used to
describe the interaction between wave packets in dispersive conservative
media, and also the interaction between orthogonally polarized components in
nonlinear optical fibres (see, e.g. \cite{Kerr,Yang1} and references
therein).

\begin{figure}[htb]
\centering \includegraphics[width=14cm]{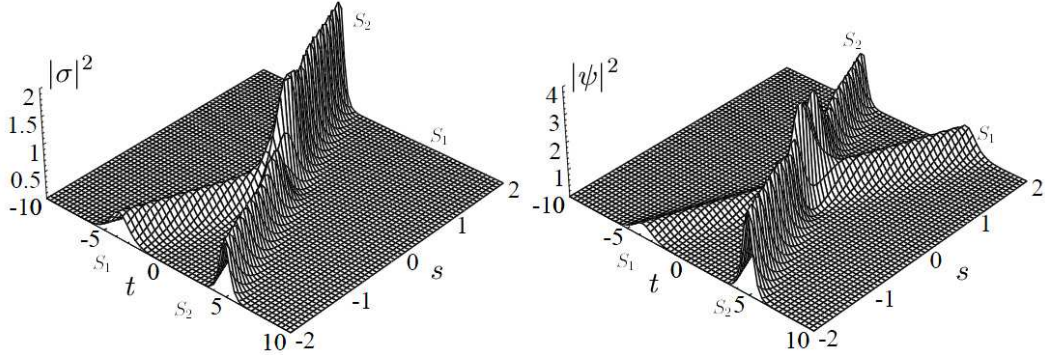}
\caption{Hypothetical market scenario including sample PDFs for volatility $|\mathcal{\protect\sigma }|^{2}$ and $|\protect\psi |^{2}$ of the Manakov 2--soliton (\protect\ref%
{ManSol}). On the left, we observe the $(s,t)-$evolution of stochastic volatility: we have a collision of two volatility component-solitons, $S_1(s,t)$ and $S_2(s,t)$, which join together into the resulting soliton $S_2(s,t)$, annihilating the $S_1(s,t)$ component in the process. On the right, we observe the $(s,t)-$evolution of option price: we have a collision of two option component-solitons, $S_1(s,t)$ and $S_2(s,t)$, which pass through each other without much change, except at the collision point. Due to symmetry of the Manakov system, volatility and option price can exchange their roles.}
\label{SolitonCollision}
\end{figure}

The simplest solution of (\ref{stochVol})--(\ref{stochPrice}), the so-called
\textit{Manakov bright 2--soliton}, has the form resembling that of (\ref%
{sech1}) and (Figure \ref{Sech1}) (see \cite{Benney,Zakharov,Hasegawa,Radhakrishnan,Agrawal,Yang,Elgin}), defined by:
\begin{equation}
\mathbf{\psi }_{\mathrm{sol}}(s,t)=2b\,\mathbf{c\,}\mathrm{sech}(2b(s+4at))\,%
\mathrm{e}^{-2\mathrm{i}(2a^{2}t+as-2b^{2}t)},  \label{ManSol}
\end{equation}%
where $\mathbf{\psi }_{\mathrm{sol}}(s,t)=\left(
\begin{array}{c}
\sigma (s,t) \\
\psi (s,t)%
\end{array}%
\right) $, $\mathbf{c}=(c_{1},c_{2})^{T}$ is a unit vector such that $|c_{1}|^{2}+|c_{2}|^{2}=1$. Real-valued parameters $a$ and $b$ are some simple functions of $(\sigma, \beta, k)$, which can be determined by either Hebbian learning of Levenberg--Marquardt algorithm. Also, shock-wave solutions similar to (\ref{tanh1}) are derived in Appendix. We can argue that in some short-time financial situations, the adaptation effect (\ref{Hebb1}) can be neglected, so our option-pricing model (\ref{stochVol})--(\ref{stochPrice}) can be reduced to the Manakov 2--soliton model (\ref{ManSol}), as depicted and explained in Figure \ref{SolitonCollision}.

More complex exact soliton solutions have been derived for the Manakov
system (\ref{stochVol})--(\ref{stochPrice}) with different procedures (see
Appendix, as well as \cite{R14,R15,R16}). For example, in \cite{R14}, using
bright one-soliton solutions (of the type of (\ref{sech1})) of the system (%
\ref{stochVol})--(\ref{stochPrice}), many physical phenomena, such as
unstable birefringence property, soliton trapping and daughter wave
('shadow') formation, were studied. Similarly, searching for modulation
instabilities and homoclinic orbits was performed in \cite{fsw00,fmmw00,wf00}%
. In particular, \emph{local bifurcations} of `wave and daughter waves' from
single-component waves have been studied in various forms of coupled
NLS--systems, including the Manakov system (see \cite{Champneys} and
references therein). Let us assume that a small volatility $\sigma -$%
component bifurcates from a pure option-price $\psi -$pulse. Thus, at the
bifurcation point, the volatility component is infinitesimally small, while
the option-price component is governed by the equation
\begin{equation*}
\partial _{ss}\psi -\psi +\psi ^{3}=0,
\end{equation*}%
whose \emph{homoclinic soliton} solution is
\begin{equation}
\psi (s)=\sqrt{2}\,\mbox{sech}\,s.  \label{upulse}
\end{equation}%
A necessary condition for a local bifurcation of a homoclinic soliton solution with
a small-amplitude volatility component from the option-price pulse (\ref%
{upulse}) is that there is a non-trivial localized solution to the
linearized problem of the $\sigma -$component. This takes the form of a
linear Schr\"{o}dinger equation
\begin{equation}
\partial _{ss}\sigma -\omega ^{2}\sigma +2\mbox{sech}^{2}s\,\sigma =0,
\label{schroedinger}
\end{equation}%
which can be solved exactly (see \cite{LaLi:77}), and for local bifurcation
we require $\sigma \rightarrow 0~~\mbox{as}~~|s|\rightarrow \pm \infty .$

As a final remark, numerical solution of the adaptive Manakov system (\ref{stochVol})--(\ref{stochPrice})--(\ref{Hebb1}), with any possible extensions, is quite straightforward, using the powerful numerical \emph{method of lines} (see Appendix in \cite{QnnBk}). Another possibility is Berger-Oliger adaptive mesh refinement when recursively numerically solving partial differential equations with wave-like solutions, using characteristic (double-null) grids (see \cite{Pretorius} and reference therein).

\subsection{Hebbian learning dynamics: analytical solution}

Regarding the Hebbian learning (\ref{Hebb1}) embedded into the Manakov
system (\ref{stochVol})--(\ref{stochPrice}), suppose e.g. that we
have $N=10$ synaptic weights (in a single neural layer), with the learning
rate $c=0.7$. The zero-mean Gaussians are defined as:
\begin{equation*}
g_{i}=\mathrm{e}^{-~\frac{t^{2}}{2\sigma _{i}}},\qquad (i=1,...,N),
\end{equation*}%
where $\{\sigma _{i}\}$ are $(-1,+1)-$random standard deviations. Using
random initial conditions, we get (by \emph{Mathematica} of \emph{Maple}
ODE-solvers) the following analytical solutions of the Hebbian learning
ODEs:
\begin{equation*}
\begin{array}{l}
w_{1}(t)=\mathrm{e}^{-t}\left[ 136485~\text{erf}(0.686579(106069~t-1))~|\psi
|^{2}+0.912318~|\psi |^{2}-0.00675663\right] , \\
w_{2}(t)=\mathrm{e}^{-t}\left[ 0.932205~\text{erf}(0.553239(16336~t-1))~|%
\psi |^{2}+0.527646~|\psi |^{2}-0.249822\right] , \\
w_{3}(t)=\mathrm{e}^{-t}\left[ 0.471627~\text{erfi}(0.477447(2.19341~t+1))~|%
\psi |^{2}-0.274787~|\psi |^{2}+0.582548\right] , \\
w_{4}(t)=\mathrm{e}^{-t}\left[ 0.52899~\text{erfi}(0.6535(117079~t+1))~|\psi
|^{2}-0.453506~|\psi |^{2}+0.0773187\right] , \\
w_{5}(t)=\mathrm{e}^{-t}\left[ 0.362728~\text{erfi}(0.324902(4.73659~t+1))~|%
\psi |^{2}-0.137812~|\psi |^{2}+0.798481\right] , \\
w_{6}(t)=\mathrm{e}^{-t}\left[ 0.523292~\text{erfi}(0.6177(131043~t+1))~|%
\psi |^{2}-0.416953~|\psi |^{2}-0.288671\right] , \\
w_{7}(t)=\mathrm{e}^{-t}\left[ 0.692907~\text{erf}(0.454319(2.42241~t-1))~|%
\psi |^{2}+0.332217~|\psi |^{2}+0.761879\right] , \\
w_{8}(t)=\mathrm{e}^{-t}\left[ 0.432141~\text{erf}(0.315332(5.02843~t-1))~|%
\psi |^{2}+0.148814~|\psi |^{2}-0.33264\right] , \\
w_{9}(t)=\mathrm{e}^{-t}\left[ 0.530916~\text{erfi}(0.673709(11016~t+1))~|%
\psi |^{2}-0.473963~|\psi |^{2}-0.17079\right] , \\
w_{10}(t)=\mathrm{e}^{-t}\left[ 0.443395~\text{erf}(0.322143(4.81806~t-1))~|%
\psi |^{2}+0.155768~|\psi |^{2}+0.49451\right] ,%
\end{array}%
\end{equation*}%
where $\text{erf}(s)$ denotes the real-valued error function, while $\text{%
erfi}(s)$ denotes the imaginary error function defined as: $\text{erf}(%
\mathrm{i}\,s)/\mathrm{i}$.
\begin{figure}[htb]
\centering \includegraphics[width=6cm]{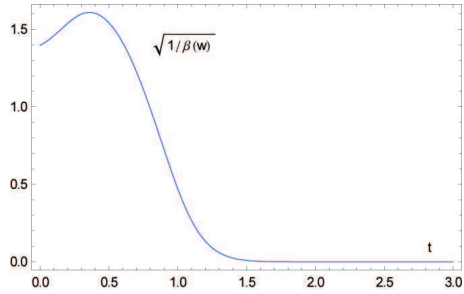}
\caption{Time plot of the quick adaptive potential term $\protect\sqrt{1/%
\protect\beta (w)}$ (as it appears in $\protect\psi _{i}(s,t),~~i=1,...,4$)
for the sample value of $\protect\psi (s,t)=0.5$.}
\label{beta}
\end{figure}

In this way, we get the alternative expression for adaptive market--heat
potential: $\beta (w)=r\sum_{i=1}^{N}w_{i}g_{i},$ with interest rate $r$
(see Figure \ref{beta}). Insertion of $\beta (w)$, including the product $%
|\sigma (s,t)||\psi (s,t)|$ calculated at time $t$ into any Manakov
solutions, gives the recursive QNN dynamics $\psi (s,t+1)$ for volatility
and option-price forecasting at time $t+1$. For example, an instant snapshot of
the adaptive bright sech-soliton $\psi _{4}(s,t)$ is given in Figure \ref%
{sechLearn}.
\begin{figure}[htb]
\centering \includegraphics[width=12cm]{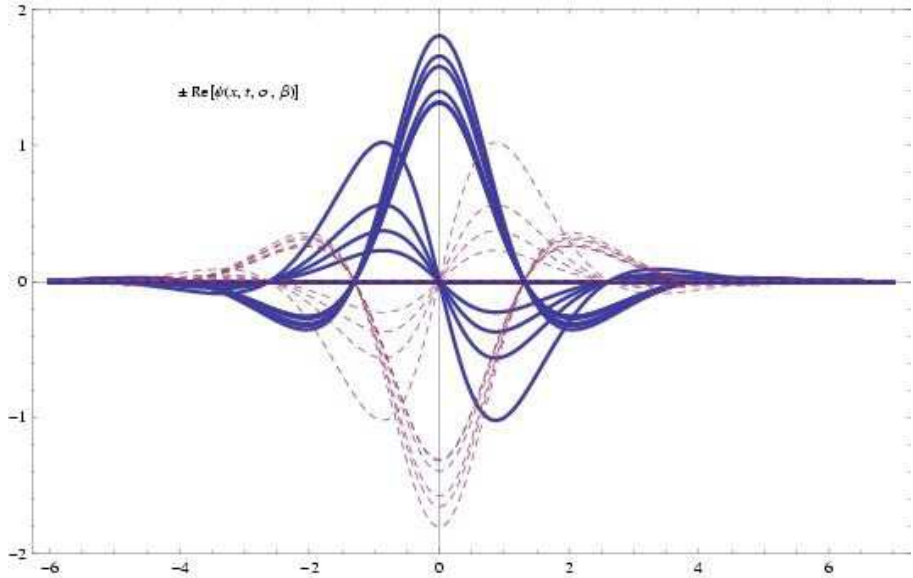}
\caption{A snapshot of the adaptive $\pm $sech-soliton $\protect\psi %
_{4}(s,t)$ with stochastic volatility $\protect\sigma _{t}$ and trained
potential $\protect\beta (w)$ calculated at a sample fixed time $t_{0}$. We
can see that due to quick learning dynamics, the whole solution is now
decaying much faster than in Figure \protect\ref{Sech1}.}
\label{sechLearn}
\end{figure}

\section{Conclusion}

I have proposed a nonlinear adaptive--wave alternative to the standard Black-Scholes option pricing model. The new model, philosophically founded on adaptive markets hypothesis \cite{Lo1,Lo2} and Elliott wave market theory \cite{Elliott1,Elliott2},
describes adaptively controlled Brownian market behavior, formally defined by adaptive NLS-equation. Four types of analytical solutions of the NLS equation are provided in terms of Jacobi elliptic functions, all starting from de Broglie's plane waves associated with the free quantum-mechanical particle. The best agreement with the Black-Scholes model shows the adaptive shock-wave NLS-solution, which can be efficiently combined with adaptive solitary-wave NLS-solution. Adjustable 'weights' of the adaptive market potential are estimated using either unsupervised Hebbian learning, or supervised Levenberg-Marquardt algorithm. For the case of stochastic volatility, it is itself represented by the wave function, so we come to the integrable Manakov system of two coupled NLS equations with the common adaptive potential, defining a bidirectional spatio-temporal associative memory machine.

As depicted in most Figures in this paper, the presented adaptive--wave model, both the single NLS-equation (\ref{nlsGen}) and the coupled NLS-system (\ref{stochVol})--(\ref{stochPrice})--(\ref{Hebb1}), which represents a bidirectional associative memory, is a spatio-temporal dynamical system of great nonlinear complexity (see \cite{ComNonlin}), much more complex then the Black-Scholes model. This makes the new wave model harder to analyze, but at the same time, its immense variety is potentially much closer to the real financial market complexity, especially at the time of economic crisis abundant in shock-waves.

Finally, close in spirit to the adaptive--wave model is the \emph{method of adaptive wavelets} in modern signal processing (see \cite{Mallet} and references therein, as well as \cite{QnnBk} for an overview), which could be used for various market dimensionality reduction, signals separation and denoising as well as optimization of discriminatory market information.

\section{Appendix: Manakov system}

\begin{figure}[htb]
\centering \includegraphics[width=16cm]{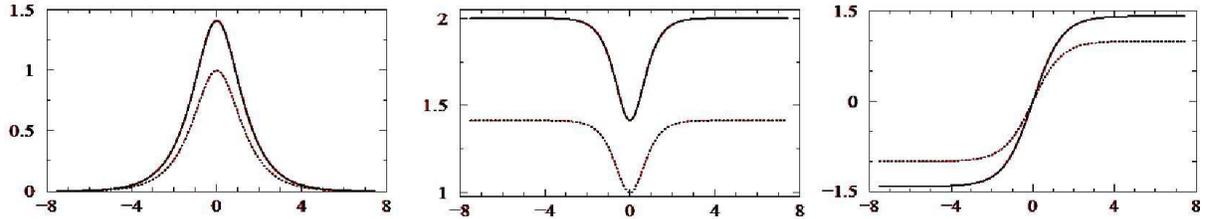}
\caption{Initial envelopes for volatility $|\protect\sigma_0|$ and
option-price $|\protect\psi_0|$ within the Manakov solitons: (left) bright
compound soliton (\protect\ref{hump}), (middle) dark compound soliton (%
\protect\ref{dark}), and (right) compound kink-shaped soliton (\protect\ref%
{kink}). These initial envelopes can be used for numerical studies of the
Manakov system and its various generalizations (modified and adapted from \protect\cite{Lazarides}).}
\label{kinkFig}
\end{figure}

Manakov's own method was based on the \textit{Lax pair representation}.\footnote{%
The Manakov system (\ref{stochVol})--(\ref{stochPrice}) has the following
Lax pair \cite{Lax} representation:
\begin{eqnarray}
\partial _{x}\phi =M\phi \ \text{\ \ \ and \ \ }\partial _{t}\phi =B\phi ,%
\text{ \ \ or \ \ }\partial _{x}B-\partial _{t}M=[M,B],\qquad \text{with}
\label{man2} \\
M(\lambda )=\left(
\begin{array}{ccc}
-\mathrm{i}\lambda & \psi _{1} & \psi _{2} \\
-\psi _{1} & \mathrm{i}\lambda & 0 \\
-\psi _{2} & 0 & \mathrm{i}\lambda%
\end{array}
\right) \qquad \text{and}  \notag \\
B(\lambda )=-\mathrm{i}\left(
\begin{array}{ccc}
2\lambda ^{2}-|\psi _{1}|^{2}-|\psi _{2}|^{2} & 2\mathrm{i}\psi _{1}\lambda
-\partial _{x}\psi _{1} & 2\mathrm{i}\psi _{2}\lambda -\partial _{x}\psi _{2}
\\
-2\mathrm{i}\psi _{1}^{\ast }\lambda -\partial _{x}\psi _{1}^{\ast } &
-2\lambda ^{2}+|\psi _{1}|^{2} & \psi _{1}^{\ast }\psi _{2} \\
-2\mathrm{i}\psi _{2}^{\ast }\lambda -\partial _{x}\psi _{2}^{\ast } & \psi
_{1}\psi _{2}^{\ast } & -2\lambda ^{2}+|\psi _{2}|^{2}%
\end{array}
\right).  \notag
\end{eqnarray}%
} Alternatively, for normalized value of the market--heat potential, $\beta=r=1$, Manakov
system allows solutions of the form:
\begin{equation}
\sigma (s,t)=\varphi (s)\,\mathrm{e}^{\mathrm{i}w_\sigma^{2}t},\qquad \psi
(s,t)=\phi (s)\,\mathrm{e}^{\mathrm{i}w_\psi^{2}t},  \label{35}
\end{equation}%
where $\varphi $, $\phi $ are real-valued functions and $w_\sigma,w_\psi$
are positive wave parameters for volatility and option-price. Substituting (%
\ref{35}) into the Manakov equations we get the ODE-system \cite{Yang1}
\begin{eqnarray}
\varphi ^{\prime \prime }(s) &=&w_\sigma^{2}\varphi (s)-[\varphi
^{2}(s)+\phi ^{2}(s)]\,\varphi (s),  \label{36} \\
\phi ^{\prime \prime }(s) &=&w_\psi^{2}\phi (s)-[\phi ^{2}(s)+\varphi
^{2}(s)]\,\phi (s).  \label{37}
\end{eqnarray}%
For $w_\sigma=w_\psi=w$, equations (\ref{36})-(\ref{37}) have a
one-parameter family of symmetric single-humped soliton solutions (see the
left part of Figure \ref{kinkFig}) given by
\begin{equation}
\varphi (s)=\pm \phi (s)=w\,\mathrm{sech}(w\,s),  \label{hump}
\end{equation}%
as well as periodic solutions:
\begin{equation}
\varphi (s)=A\cos (Bs)\qquad \text{and}\qquad \phi (s)=A\sin (Bs),
\label{per1}
\end{equation}%
where $A=\sqrt{w^{2}+B^{2}}$ (with $B$ an arbitrary parameter). For $0<w<1$
there is also another, in general asymmetric, one-parameter family of
solutions for each fixed $w$ \cite{Yang1}
\begin{eqnarray}
\varphi (s) &=&{\ }\sqrt{2(1-w^{2})}\cosh {(}ws{)}/{\kappa ,}  \label{43} \\
\phi (s) &=&{\ -}w\sqrt{2(1-w^{2})}\sinh {(}s-s_{0}{)}/{\kappa },\qquad
\text{where}  \notag \\
\kappa &=&\cosh (s)\cosh (w\,s)-w\sinh (s)\sinh (w\,s),  \notag
\end{eqnarray}%
in which $\varphi $ is symmetric and $\phi $ antisymmetric.

On the other hand, for negative values of the potential $\beta $, the
Manakov equations accept dark soliton solutions of the form \cite{Lazarides}
\begin{equation}
\sigma (s,t)=\psi (s,t)=k\left[ \tanh (ks)-\mathrm{i}\right] \mathrm{e}^{%
\mathrm{i}(ks-5k^{2}t)},  \label{dark}
\end{equation}%
which are localized dips on a finite-amplitude background wave (see the
middle part of Figure \ref{kinkFig}). In this very interesting case,
volatility and option-price fields are coupled together, forming a dark
compound soliton. Note that their respective relative amplitudes are
controlled by the corresponding nonlinearities and frequency. For $\beta =-1$
the Manakov equations alow also solutions of the form:
\begin{equation}
\sigma (s,t)=\varphi (s)\,\mathrm{e}^{-\mathrm{i}w_\sigma^{2}t},\qquad
\,\,\psi (s,t)=\phi (s)\,\mathrm{e}^{-\mathrm{i}w_\psi^{2}t}.  \label{44}
\end{equation}%
Introducing (\ref{44}) into the Manakov equations, we get the ODE-system:
\begin{eqnarray}
\varphi ^{\prime \prime }(s) &=&[\varphi ^{2}(s)+\phi ^{2}(s)]\,\varphi
(s)-w_\sigma^{2}\varphi (s),  \label{45} \\
\phi ^{\prime \prime }(s) &=&[\phi ^{2}(s)+\varphi ^{2}(s)]\,\phi
(s)-w_\psi^{2}\phi (s),  \label{46}
\end{eqnarray}%
which, for $w_\sigma=w_\psi=w$, allow for kink-shaped localized soliton
solutions (see the right part of Figure \ref{kinkFig}) given by \cite%
{Lazarides}
\begin{equation}
\varphi (s)=\pm \phi (s)=({w}/\sqrt{2})\tanh ({w}\,s/\sqrt{2}),  \label{kink}
\end{equation}%
as well as periodic solutions (\ref{per1}). Inserting (\ref{kinkFig}) back
into (\ref{44}) gives the double-kink solution for the Manakov system:
\begin{equation}
\sigma (s,t)=\pm({w}/\sqrt{2})\tanh ({w}\,s/\sqrt{2})\,\mathrm{e}^{-\mathrm{i%
}w^{2}t},\qquad \,\,\psi (s,t)=\pm({w}/\sqrt{2})\tanh ({w}\,s/\sqrt{2})%
\mathrm{e}^{-\mathrm{i}w^{2}t}.  \label{2kink}
\end{equation}

\end{document}